%% file: main.tex
\documentclass[sigconf]{acmart}

\AtBeginDocument{%
  }

\setcopyright{acmlicensed}
\copyrightyear{2024}
\acmYear{2024}
\acmDOI{XXXXXXX.XXXXXXX}

\usepackage{adjustbox}
\usepackage{threeparttable}
\usepackage{multirow}
\usepackage{multicol}
\usepackage{subfigure}

\usepackage{tcolorbox}

\newcommand{\finding}[1]{ \begin{tcolorbox}#1\end{tcolorbox}}

\begin{document}

\title{RecSys Arena: Pair-wise Recommender System Evaluation with Large Language Models}

\author{Zhuo Wu}
\email{wuzhuo@tju.edu.cn}
\affiliation{%
  \institution{School of New Media and Communication, Tianjin University}
  \city{Tianjin}
  \country{China}
}

\author{Qinglin Jia}
\email{jiaqinglin2@huawei.com}
\affiliation{%
  \institution{Huawei Noah's Ark Lab}
  \city{Beijing}
  \country{China}}

\author{Chuhan Wu}
\email{wuchuhan15@gmail.com}
\affiliation{%
  \institution{Huawei Noah's Ark Lab}
  \city{Beijing}
  \country{China}
}

\author{Zhaocheng Du}
\email{zhaochengdu@huawei.com}
\affiliation{%
  \institution{Huawei Noah's Ark Lab}
  \city{Beijing}
  \country{China}
}

\author{Shuai Wang}
\email{wangshuai231@huawei.com}
\affiliation{%
  \institution{Huawei Noah's Ark Lab}
  \city{Beijing}
  \country{China}
}

\author{Zan Wang}
\email{wangzan@tju.edu.cn}
\affiliation{%
  \institution{School of New Media and Communication, Tianjin University; College of Intelligence and Computing, Tianjin University}
  \city{Tianjin}
  \country{China}
}

\author{Zhenhua Dong}
\email{dongzhenhua@huawei.com}
\affiliation{%
  \institution{Huawei Noah's Ark Lab}
  \city{Beijing}
  \country{China}
}

\renewcommand{\shortauthors}{Wu et al.}

\begin{abstract}

Evaluating the quality of recommender systems is critical for algorithm design and optimization.
Most evaluation methods are computed based on offline metrics for quick algorithm evolution, since online experiments are usually risky and time-consuming.
However, offline evaluation usually cannot fully reflect users' preference for the outcome of different recommendation algorithms, and the results may not be consistent with online A/B test.
Moreover, many offline metrics such as AUC do not offer sufficient information for comparing the subtle differences between two competitive recommender systems in different aspects, which may lead to substantial performance differences in long-term online serving.
Fortunately, due to the strong commonsense knowledge and role-play capability of large language models (LLMs), it is possible to obtain simulated user feedback on offline recommendation results.
Motivated by the idea of LLM Chatbot Arena, in this paper we present the idea of RecSys Arena, where the recommendation results given by two different recommender systems in each session are evaluated by an LLM judger to obtain fine-grained evaluation feedback.
More specifically, for each sample we use LLM to generate a user profile description based on user behavior history or off-the-shelf profile features, which is used to guide LLM to play the role of this user and evaluate the relative preference for two recommendation results generated by different models.
Through extensive experiments on two recommendation datasets in different scenarios, we demonstrate that many different LLMs not only provide general evaluation results that are highly consistent with canonical offline metrics, but also provide rich insight in many subjective aspects.
Moreover, it can better distinguish different algorithms with comparable performance in terms of AUC and nDCG.
Our codes are publicly available at \url{https://github.com/anonyProjects/RecSys-Arena}.

\end{abstract}
\acmArticleType{Review}
\acmCodeLink{https://github.com/borisveytsman/acmart}
\acmDataLink{htps://zenodo.org/link}
\acmContributions{BT and GKMT designed the study; LT, VB, and AP
  conducted the experiments, BR, HC, CP and JS analyzed the results,
  JPK developed analytical predictions, all authors participated in
  writing the manuscript.}

\maketitle

\section{INTRODUCTION}
\label{sec:intro}

\input{content/intro}

\section{METHODOLOGY}
\label{sec:method}
\input{content/method}

\section{EXPERIMENTAL SETUP}
\label{sec:evaluation}
\input{content/evaluation}

\section{EXPERIMENTAL RESULTS}
\label{sec:result}
\input{content/result}

\section{RELATED WORK}
\label{sec:related}
\input{content/relatedWork}

\section{CONCLUSION}
In this paper, we propose a practical LLM-based pair-wise evaluation method for recommender systems. Our experiments primarily investigate the feasibility of using LLMs for comparative evaluation in this context. We assess both the overall effectiveness and the sub-dimensional effectiveness of LLMs in the pair-wise evaluation task, examining their performance across various LLMs. Furthermore, we find that the LLM-based pair-wise evaluation method not only produces results that align with the trends of offline metrics but also offers improved discrimination, capturing finer distinctions between different recommendation models.

\bibliographystyle{ACM-Reference-Format}
\bibliography{main}

\end{document}

%% file: content/intro.tex
Accurate and comprehensive evaluation of recommendation algorithms is essential in practical recommender system design and optimization~\cite{zangerle2022evaluating, bauer2024exploring}.
However, recommender system evaluation is very challenging due to the complexity of user feedback and rapid shift of data distribution. 
Although online experiments such as A/B testing can give direct assessments about the overall performance of different recommendation algorithms, it is relatively time-consuming to accumulate sufficient user behaviors to obtain confident results.
Moreover, many mainstream metrics such as click-through rate (CTR) do not fully reflect user satisfaction and long-term user experience.

To help system designers optimize their algorithms, researchers usually evaluate the performance of recommender systems based on offline user behavior logs to obtain preliminary assessments~\cite{beel2013research}.
For example, metrics such as AUC and nDCG are widely used in different domains to indicate the ranking quality of recommender systems~\cite{zhu2022bars}.
In addition, researchers devise various metrics to quantitatively measure the behaviors of models in different aspects, such as coverage, diversity, novelty and serendipity~\cite{zangerle2022evaluating}.
However, these offline metrics may not well reflect the user responses to recommendation results, thereby may not be consistent with the results of online experiments.
In addition, many ranking metrics such as AUC are not sufficiently sensitive to distinguish the real quality of different recommender systems, which may not provide valuable reference for picking promising candidate algorithm for online experiments.

\begin{figure*}[t]
    \centering
    \includegraphics[width=0.8\textwidth]{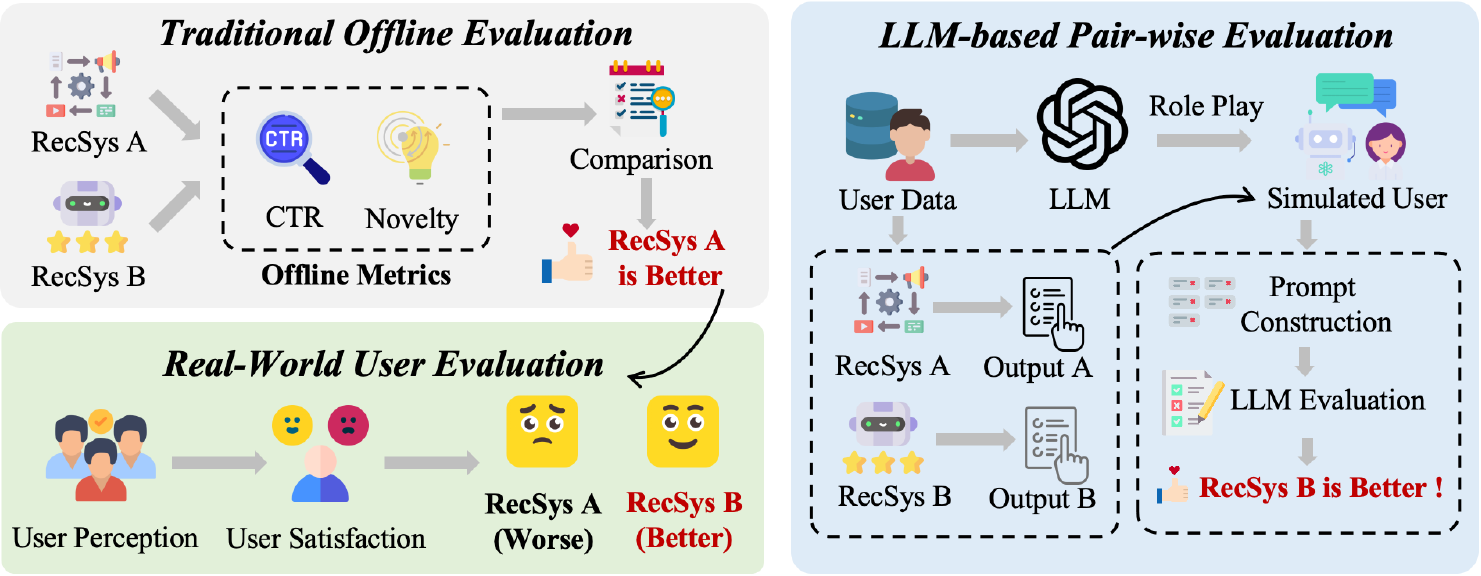}
    \caption{The differences among traditional offline evaluation, real-world user evaluation, and LLM-based pair-wise evaluation \label{fig:overview}}
\end{figure*}


In recent years, researchers explore the use of large language models (LLMs) in recommender system evaluation due to their rich general knowledge memorization and human-oriented behavior alignment~\cite{wang2023rethinking, zhang2024large}. 
For example, zhang et al.~\cite{zhang2024large} proposed that certain zero-shot LLMs can achieve comparable or even better evaluation accuracy compared to traditional methods in the task of assessing recommendation explanation quality. In addition, they claimed that using the voting results of multiple LLMs can improve the accuracy of evaluations.
Wang et al.~\cite{wang2023rethinking} proposed utilizing the role-play capability of LLMs and using them as user simulators to evaluate conversational recommender systems (CRSs).
These methods usually rely on LLMs to generate direct assessments of the recommendation results, which may not reflect the relative quality difference between different recommender systems.

Inspired by the relative evaluation methods of LLM such as Chatbot Arena~\cite{chiang2024chatbot} and AlpacaEval~\cite{dubois2024length}, 
we propose a practical LLM-based pair-wise evaluation framework named RecSys Arena, which aims to evaluate the relative performance of recommendation methods on each sample.
As demonstrated by the existing work~\cite{dai2023uncovering}, in recommendation task, LLM is good at pair-wise ranking while less good at point-wise ranking.
In the relative evaluation task, LLMs can simultaneously access information from two recommendation results, facilitating a more granular comparative analysis and uncovering subtle differences to assess their alignment with user preferences.

Figure~\ref{fig:overview} presents the overview of our approach RecSys Arena.
To tackle the limitation of offline evaluations in accurately reflecting user perceptions, we utilize LLMs to simulate users.
More specifically, we extract user information from various data sources, including behavioral history and existing profile features. This data is then used to construct a detailed user profile description. By simulating the role of the user, the LLM can generate personalized evaluation.
Next, to facilitate pair-wise evaluation, we provide the LLM with the recommendation result lists generated by two different recommender systems when constructing the prompts.
Compared to absolute evaluation, relative evaluation offers more contrast information, allowing the LLM to perform a finer-grained assessment of the recommendation results. This approach enhances the LLM's ability to distinguish subtle differences in recommendation results.
At the same time, compared to online evaluations, the LLM-based pair-wise evaluation method offers greater feasibility and efficiency.
This method allows for rapid testing of different recommendation scenarios, enabling researchers to analyze large datasets and assess various recommendation model performances. Furthermore, by leveraging LLMs, the evaluation process can be conducted at scale, providing a more comprehensive understanding of user preferences while reducing the time and resources typically required for online evaluations.
In summary, we leverage the human-like and role-play capabilities of LLMs to conduct pair-wise evaluations of recommendation results, assessing which of the two recommender systems performs better based on an understanding of the user's personal attributes and preferences.
Moreover, LLMs are pre-trained on vast data corpora in a self-supervised manner, allowing them to capture extensive domain knowledge. This exposure helps them learn intricate patterns and contextual cues, enhancing their reasoning abilities~\cite{brown2020language}.
Additionally, with billions of parameters fine-tuned during training, these models can effectively encode and recall information, facilitating reasoning processes.
For example, for categories of items that do not appear in the historical interactions, the LLM will conduct a potential inferential analysis of whether the user might be interested in the item based on personal attributes or other information.

We conducted an experimental study to demonstrate the effectiveness of the LLM-based method in pair-wise evaluation of recommender system performance.
In our study, we considered different types of recommendation models, including factorization machines~\cite{rendle2010factorization}, ID-based recommendation model~\cite{guo2017deepfm}, content-based recommendation model~\cite{wu2019neural, an2019neural}, sequence recommendation model~\cite{kang2018self}, and graph neural network-based recommendation model~\cite{he2020lightgcn, wang2019neural, mao2021ultragcn, hamilton2017inductive}.
We used two public content recommendation datasets (i.e. MovieLens~\cite{harper2015movielens} and MIND~\cite{wu2020mind}) for evaluation.
In our study, we considered both open-source and closed-source LLMs across various sizes, ranging from 8 billion to 236 billion parameters.
Additionally, we designed six aspects for evaluating the quality of recommendation results from the user's perspective.

Our study makes the following findings:
\begin{itemize}
    \item Large language models, leveraging their reasoning capabilities, world knowledge, text generation capabilities, and role-play capabilities, can generate reasonable pair-wise evaluation results. Moreover, when comparing two recommendation models, these results align with the trends observed in offline metrics, such as AUC and Diversity.
    \item Different large language models exhibit varying effectiveness in the task of recommendation quality evaluation, with larger models generally performing better.
    \item Pair-wise evaluation based on large language models offers a more nuanced distinction between two different recommendation models with similar performance in terms of AUC and nDCG. RecSys Arena can uncover subtler differences in recommendation results that existing offline metrics might overlook.
\end{itemize}


%% file: content/method.tex
In this article, we propose a novel and practical approach , called RecSys Arena, to utilize the LLM to conduct pair-wise evaluations of the two recommender systems.
The entire evaluation process consists of four steps.
First, two recommender systems generate recommendation result lists for the same user. Next, user information, recommendation results, and descriptions of the evaluation aspects are integrated to construct a prompt. The prompt is then input into a large language model to obtain qualitative analyses and quantitative comparison results for each evaluation aspect. Finally, an evaluation report is produced.

\subsection{Problem Formulation}

We use $U = (u_1,u_2,...,u_{|U|})$ to denote the set of users in a recommender system (RS). 
Input to the RS includes the user's personal attribute information $S$ and viewing histories $H$ of users respectively.
The RS recommend multiple items to each user $u$, which are defined as $I_u = (i_1,i_2,...,i_{|I_u|})$.

Given two RSs $R_A$ and $R_B$, we use $I_{R_A}$ and $I_{R_B}$ to represent the corresponding recommendation result lists generated by systems $R_A$ and $R_B$, respectively. 
Let $f(\cdot)$ represent the evaluation method.
The pair-wise evaluation results of systems $R_A$ and $R_B$, as provided by the LLM, can be expressed as:
\begin{equation}
f_{LLM}(U,R_A,R_B) = LLM(S_U, H_U, I_{R_A},I_{R_B},\mathcal{P})
\end{equation}
where $\mathcal{P}$ denotes the prompt template. 

The primary goal of pair-wise LLM-based evaluation is to 1) measure, from the user's perspective, which recommender system, $R_A$ or $R_B$, has better overall performance for the same user. Note that the overall performance refers to the comparison results that take into account multiple evaluation dimensions (to be introduce in ~\ref{sec:evaluation_dimensions}).
Along with the measurable overall performance, 2) LLM-based evaluation method also reports a detailed, interpretable qualitative analysis explaining the evaluation reasons for each dimension, which could facilitate the developer to further make targeted improvements to the recommender system.

\subsection{Evaluation Dimensions}
\label{sec:evaluation_dimensions}

To address the issue that existing offline evaluation metrics cannot evaluate the quality of recommendation results from the user's perception, we primarily focus on user experience when designing the evaluation dimensions.
We assume that users of the recommender system serve as the most accurate evaluators of the recommendation results.
We consider both mainstream dimensions of concern (e.g., accuracy, satisfaction) and dimensions that traditional offline metrics cannot evaluate (e.g., inspiring content, positive impact).
Therefore, for the paired recommendation result $I_{R_A}$ and $I_{R_B}$, the LLM is asked to give a comparative evaluation result from the following 6 aspects:

\textbf{Accuracy:} This recommendation result list aligns well with my interests.

\textbf{Satisfaction:} I am satisfied with the recommendation results provided by this recommender system.

\textbf{Inspiration:} The recommended items inspire me to think, promote further exploration, and enhanced my willingness to interact with the recommendation platform.

\textbf{Content Quality:} The recommended items are of high quality.

\textbf{Transparency:} The recommendation results are associated with one of my personal information or an interaction history, and it is evident which feature is relevant.

\textbf{Impact on users:} The impact of this recommendation result list on me is positive.



\subsection{Prompt Construction}

\begin{figure}[t]
    \centering
    \includegraphics[width=0.5\textwidth]{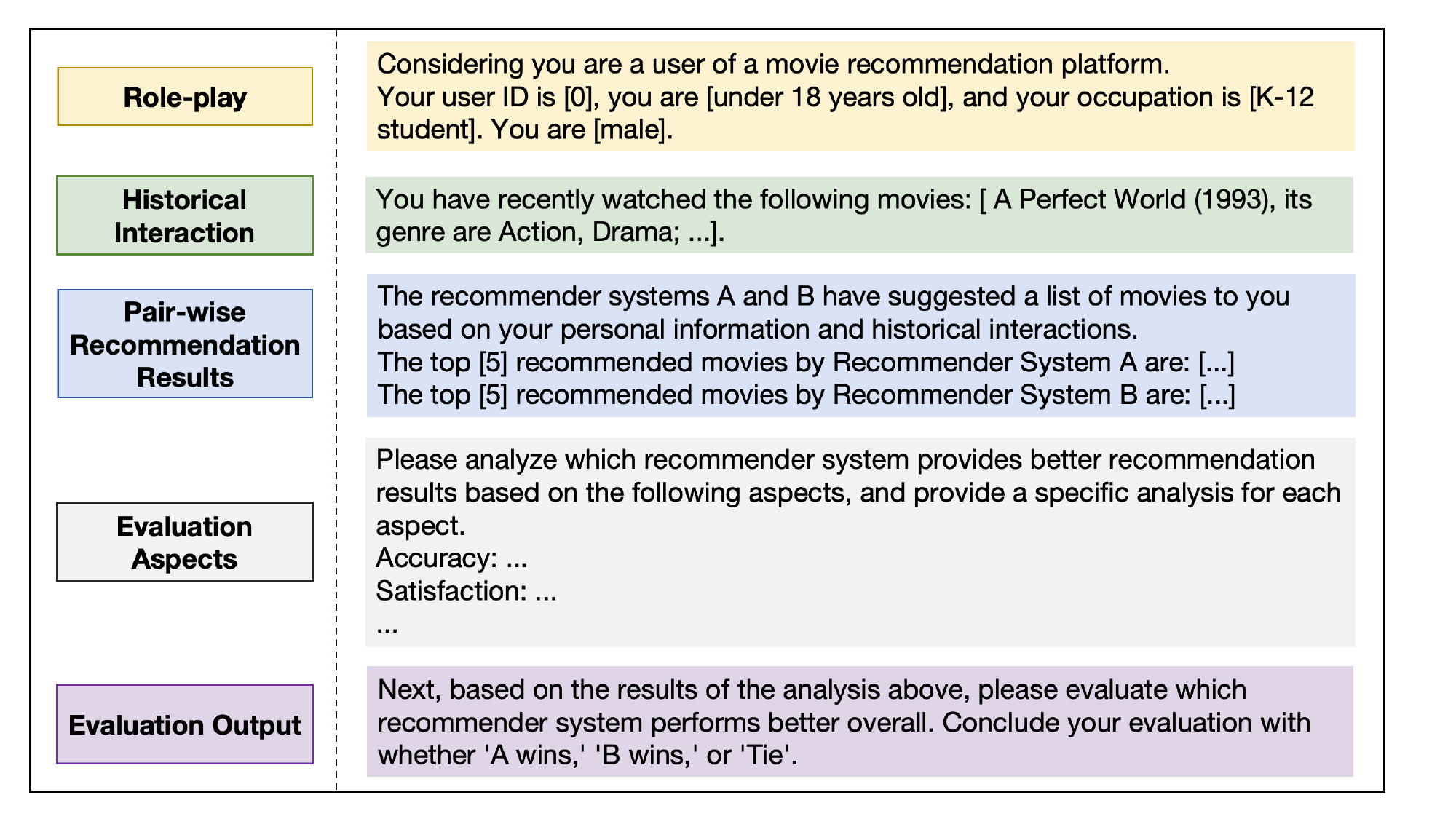}
    \caption{The outline of evaluation prompt template applied in our study \label{fig:prompt-template}}
\end{figure}

In the section, we mainly introduce how we construct the prompt $\mathcal{P}$. The prompt is designed to guide LLM to relatively evaluate the quality of pair-wise recommendation results from specific aspects, based on the user profile and viewing histories.
As shown in Figure~\ref{fig:prompt-template}, the prompt $\mathcal{P}$ includes five key components.

In the first part of the prompt, we leverage the role-play capability of LLMs to facilitate a personalized evaluation of recommendation results. To do this, we provide the LLM with the user's personal attribute information, including age, occupation, gender, and other relevant details.
The second part of the prompt consists of the user's viewing history, such as information on movies they have watched or news they have clicked on. This content is included to allow the LLM to perceive the user's preferences, enabling it to conduct subsequent evaluations from the user's perspective.
Please note that the MIND dataset does not contain any personal user information. Instead, we provide the historical records of news articles that users have browsed, allowing the LLM to understand the user profile. This approach also helps the LLM gain insights into user preferences and behaviors.
Next, the recommendation results from the two systems, $R_A$ and $R_B$, are presented to the LLM via the prompt. These recommendation results will include specific item information, such as the titles and genres of the movies.
In the evaluation section of the prompt, we list the descriptions of each evaluation dimension to assist the LLM in understanding the specific content that requires assessment for each dimension. Additionally, the evaluation dimensions in this prompt template can be dynamically adjusted, further enhancing the scalability of the evaluation framework. This flexibility allows researchers to tailor the evaluation criteria to suit different contexts and objectives, making it applicable across a wide range of scenarios.
The main objective of this section is to allow the LLM to make evaluative judgments based on its analysis of each evaluation aspect. We aim to guide the LLM through a step-by-step thought process, similar to the Chain-of-Thought~\cite{wei2022chain}. This method encourages deeper reasoning and enhances the quality of the evaluation by building on prior insights.
Finally, the LLM is asked to output the qualitative analysis for each dimension, along with an overall comparative evaluation of the pair-wise recommendation results from systems $R_A$ and $R_B$.

\subsection{LLM Evaluator Construction}
We utilize pre-trained LLMs to provide comparative evaluations for paired recommender systems.
The LLM receives the user's personal attribute information, viewing histories, and recommendation results from the two recommender systems $R_A$ and $R_B$ under test, accompanied by the prompt $\mathcal{P}$ to describe the evaluation instruction.
LLMs trained on massive corpora of unlabelled data possess a wealth of general knowledge, which aids them in understanding recommended items, such as movies.
The reason for their strong power can be concluded as they do not need task-specific training data and can be pre-trained on tremendous in-the-wild data in a self-supervised manner (a.k.a. pre-training), so that sufficient domain knowledge can be captured~\cite{radford2018improving, devlin2018bert, brown2020language}.

Previous research on evaluation based on LLMs has mostly involved absolute evaluation~\cite{zhang2024large, kocmi2023large}, where the LLM assigns scores to specified evaluation content. 
Our approach differs from previous studies in that we ask the LLM to conduct a comparative evaluation of two recommendation results, thereby providing a relative assessment.
LLMs perform better on pair-wise tasks~\cite{dai2023uncovering}.
On one hand, using relative evaluation allows the LLM to simultaneously access information from two recommendation results, facilitating a more nuanced comparative analysis. This enables the LLM to uncover subtle differences and assess how well each result aligns with user preferences. On the other hand, absolute scoring evaluations often provide limited context, making it challenging for the model to identify and distinguish between the merits of individual recommendations. By leveraging relative evaluation, we enhance the LLM's capacity to perform finer-grained assessments, ultimately leading to more accurate and personalized recommendations.

We conduct a statistical analysis of the evaluation results generated by the LLM.
To measure the degree of victory between the two models more precisely, we designed the quantile $\mathcal{Q}$ metric.
Specifically, we calculate the quantile $\mathcal{Q}$ using the following formula:
\begin{equation}
 \mathcal{Q} = \frac{(N_{win} + N_{tie})}{(N_{lose} + N_{tie})}
\end{equation}
where $N_{win}$ denotes the number of samples in the test set where the RS is deemed to have won, $N_{tie}$ indicates the number of samples where the RSs tied, and $N_{lose}$ represents the number of samples in which the RS lost.
A larger value of $\mathcal{Q}$ indicates a greater degree of victory for the RS.


%% file: content/evaluation.tex
\subsection{Research Questions}

The goal of our study is to investigate the effectiveness of large language models on recommendation performance evaluation.
To this end, we propose to answer two main research questions.

RQ1: What is the overall performance of the LLM-based evaluation method for recommender systems?
As the very first RQ, we aim at investigating the feasibility of using LLMs for comparative evaluation of the overall effectiveness of two recommender systems. Additionally, we assess the effectiveness of the LLM-based evaluation by determining whether it aligns with offline usage prediction metrics (i.e., AUC).

RQ2: What is the performance of the LLM-based evaluation method for recommender systems in different evaluation sub-dimensions?
Due to the absence of subjective user labels (e.g., user satisfaction, whether the impact on users is negative or positive, etc.) in the datasets, we are unable to directly assess the effectiveness of LLMs in evaluating sub-dimensions. We selected the dimensions of inspiration and transparency, and designed an indirect meta evaluation strategy.
First, among the currently evaluable offline metrics, there is overlap in the content assessed for diversity and inspiration. Therefore, we considered whether the evaluation results for inspiration could align with the offline metrics of diversity.
Second, we assessed whether LLMs could correctly evaluate transparency by observing whether LLMs could capture differences in feature selection during training.


\subsection{Evaluation Metrics}
We can initially assess the effectiveness of evaluation based on LLMs by determining whether the results provided by the LLMs align with offline metrics.
We consider the three popular metrics: AUC (Area under the Curve)~\cite{ling2003auc}, nDCG$@k$ (Normalized Discounted Cumulative Gain for the top $k$ recommendations)~\cite{jarvelin2002cumulated}, URD (User Recommendation Diversity)~\cite{qin2013promoting}. The first two metrics measure recommendation accuracy, while the last metric measure recommendation diversity.

\textbf{AUC}~\cite{ling2003auc} measures the area under the receiver operating characteristic curve, plotting the true positive rate against the false positive rate. It is widely used to evaluate recommender systems, indicating how often relevant items are ranked higher than irrelevant ones. A higher AUC reflects better recommendation accuracy.

\textbf{nDCG$@k$}~\cite{jarvelin2002cumulated} is determined using the formula $nDCG@k = \frac{DCG@k}{IDCG@k}$. Here, $DCG@k$ is calculated as $DCG@k = \sum_{i=1}^{k} \frac{2^{p(i)} - 1}{log_2(i+1)}$, where $p(i)$ represents the preference score of the user $u$ for the $i^{th}$ item in the generated recommendation list. $IDCG@k$ represents the DCG value of the ideal $k$-item recommendation list. 

\textbf{URD}~\cite{qin2013promoting} measures the degree of recommendation diversity. It is calculated based on the intra-list similarity~\cite{ziegler2005improving}, using the formula: $URD(rl) = 1 - \frac{2}{n(n-1)} \sum_{i_a \in rl} \sum_{i_b \neq i_a \in rl} Sim(i_a, i_b)$ represents the similarity between item $i_a$ and item $i_b$.

\subsection{Recommender Systems and Datasets}

In this study, we use two content recommendation datasets (i.e., Movielens and MIND) released by the previous study~\cite{harper2015movielens, wu2020mind} as our evaluation datasets, which have been widely used in the existing studies~\cite{lin2022improving, wu2019neural, an2019neural, xie2022contrastive, krichene2020sampled}.
More specifically, 
MovieLens~\cite{harper2015movielens} comprises data from 1 million movie ratings provided by 6,040 users across 3,883 movies. Within this dataset, user's attributes include gender, age, and occupation.
MIND~\cite{wu2020mind} is a news dataset and was collected from the user behavior logs of Microsoft News. The dataset contains one million users who had at least five news click records during six weeks from October 12 to November 22, 2019. It contains 161,013 news and 24,155,470 reading records.


In our study, we use 5 categories of recommender systems as the subject of the evaluation in total.
We considered different types of recommendation models under test: Factorization Machines, the content-based model, the graph-based model, the sequence recommendation model, and the ID-based model.
We describe the specific information of the recommendation models as follow:
\begin{itemize}
    \item FM. Rendle et al. propose factorization machine (FM)~\cite{rendle2010factorization}, which improve upon logistic regression models by addressing the challenge of training model parameters in sparse data scenarios. Furthermore, FMs incorporate second-order feature interactions, compensating for the limited expressive power of logistic regression.
    \item NRMS. NRMS~\cite{wu2019neural} is a representative of content-based recommendation models that employs multi-head self-attention mechanisms for encoding content, such as news titles.
    \item LightGCN. LightGCN~\cite{he2020lightgcn} is a model that simplifies Graph Convolutional Networks (GCNs) by focusing solely on the core component of neighborhood aggregation for collaborative filtering.
    \item SASRec. SASRec~\cite{kang2018self} is a sequence recommendation model based on self-attention mechanisms.
    \item DeepFM. DeepFM~\cite{guo2017deepfm} is an extension of Wide\&Deep that synergistically integrates factorization machines for recommendation and deep learning for feature learning, emphasizing both low- and high-order feature interactions.
\end{itemize}

\input{forms/dataset-model}
Table~\ref{tab:model_information} lists the values of the evaluation metrics (i.e., AUC and nDCG@5) for the above five recommendation models, respectively on the Movielens and MIND datasets.

\subsection{LLMs}

We conduct experiments with three LLMs, including the open-source LLM and proprietary LLM.
\begin{itemize}
    \item GPT-4o~\cite{gpt-4o/url}. GPT-4o is a proprietary LLM released by OpenAI. With targeted optimizations, GPT-4o delivers superior performance in generating accurate and contextually relevant responses, benefiting from refined training techniques and updates. 
    \item DeepSeek-V2.5~\cite{deepseekv2}. DeepSeek-V2.5 is a strong Mixture-of-Experts (MoE) langage model that excels in writing and instruction-following. It comprises 236B total parameters.
    \item Llama3.1-8B-Instruct~\cite{llama3/url}. LLaMA3-8b-Instruct is fine-tuned specifically to follow and execute user instructions more accurately, making it better at handling tasks that involve clear directives or specific commands.
\end{itemize}

To make the output as deterministic as possible, we set temperature=0 when calling the API.

%% file: forms/dataset-model.tex
\begin{table}[t!]
\centering
\caption{The recommendation accuracy of recommender systems in terms of AUC and nDCG@5  \label{tab:model_information}}
\begin{adjustbox}{width=0.4\textwidth,center}
\begin{threeparttable}
\begin{tabular}{c|cc|cc}
\toprule
    \multirow{2}{*}{RS}     & \multicolumn{2}{c|}{MovieLens} & \multicolumn{2}{c}{MIND} \\ \cline{2-5}
         & AUC    & nDCG@5      & AUC        & nDCG@5        \\ \hline
FM       & 0.5701   & 0.0557       &      0.4857      &    0.1966         \\ 
NRMS     & 0.7521   & 0.1216       &     0.5004       &    0.2242         \\ 
LightGCN & 0.6824   & 0.1101       &     0.4990       &     0.2197        \\ 
SASRec   & 0.6772   & 0.1086       &     0.4985       &     0.2190        \\ 
DeepFM   & 0.6146   & 0.0970       &      0.4956      & 0.2183\\
\bottomrule
\end{tabular}
\end{threeparttable}
\end{adjustbox}
\end{table}

%% file: content/result.tex
\subsection{RQ1: The Overall Performance of Evaluation}

\input{forms/rq1-overview-movie}

Our main results are displayed in Table~\ref{tab:rq1-movie}.
Specifically, we conducted pair-wise evaluations by separately comparing the ID-based recommendation model (i.e., DeepFM), content-based recommendation model (i.e., NRMS), sequential recommendation model (i.e., SASRec), and graph network-based recommendation model (i.e., LightGCN) with FM.
First, we present the proportions of "\textbf{Win}", "\textbf{Tie}", and "\textbf{Lose}" for each RS relative to FM in the evaluation results provided by the LLM.
Then, in Table~\ref{tab:rq1-movie}, we calculated the quantile $\mathcal{Q}$ (i.e., $ ( N_{win} + N_{tie} ) / ( N_{lose} + N_{tie} ) $) shown in Column "Q" and listed the rankings of the quantiles $\mathcal{Q}$ in Column "Rank".

To investigate the overall effectiveness of LLMs in evaluating recommendation quality, we use the AUC from traditional metrics as a reference. Specifically, we examine whether the ranking results of recommendation quality provided by LLMs are consistent with the ranking results based on traditional accuracy metric (i.e., AUC). 


\textbf{Impact of different LLMs.}
In our experiments, we employed three LLMs: GPT-4o, DeepSeek-V2.5-236B, and Llama3.1-8B-Instruct, representing closed-source large language models, large-sized open-source large language models, and smaller-sized open-source large language models, respectively.
Then, we examined differences across various LLMs.
The ranking results provided by GPT-4o and DeepSeek-V2.5-236B are consistent with traditional ranking results (as shown in Table~\ref{tab:model_information}). 
However, the ranking results provided by Llama3.1-8B-Instruct show some discrepancies.
Moreover, in the comparative evaluation results provided by Llama3.1-8B-Instruct, the likelihood of the tested models being tied is higher. This indicates that Llama3.1-8B-Instruct's evaluation criteria may be more lenient, potentially leading to a higher incidence of similar performance scores among different models.
Larger LLMs are better equipped to evaluate recommendation quality.

Additionally, to further analyze the correlation between the evaluation results derived from the LLM and the offline metrics, we calculated the Pearson Correlation~\cite{cohen2009pearson} between the AUC and the quantiles $\mathcal{Q}$, as shown in Table~\ref{tab:pearson}.
Table~\ref{tab:pearson} indicates that the evaluation results generated by GPT-4o and DeepSeek-V2.5 show a strong correlation with the offline metric AUC; however, the P-values are both greater than 0.01, suggesting that the correlation is not significant. In contrast, the evaluation results generated by Llama3.1-8B exhibit a moderate correlation with the offline metric AUC.

\input{forms/pearson}

\textbf{Impact of different Datasets.}
Furthermore, we considered the datasets conducting different tasks, including movies and news recommendations.
Compared to the Movielens dataset, the MIND dataset provides more textual content, including news titles and abstracts, which is more conducive to the LLM's understanding of items.
From the horizontal comparison of the results from both datasets, the LLM maintains a consistent win-loss ratio for evaluation of ID-based recommendation models, sequential recommendation models, and graph-based recommendation models. However, as shown in Table~\ref{tab:rq1-movie}, content-based recommendation models perform better on the MIND dataset, indicating that different types of models have different applicable scenarios. The evaluation results can guide the selection of models in different scenarios.

\begin{figure}[t]
    \centering
    \includegraphics[width=0.5\textwidth]{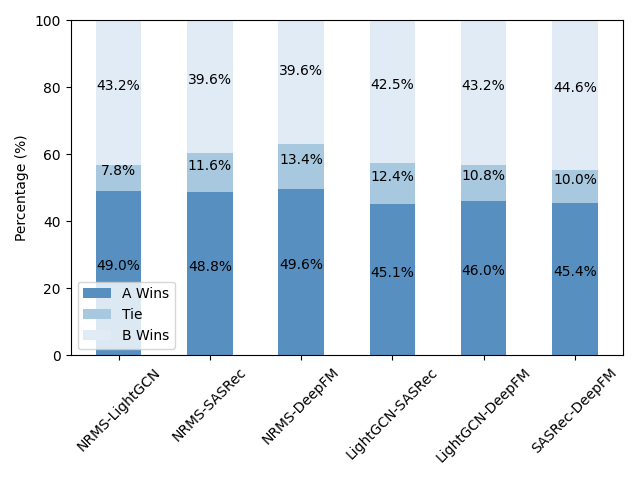}
    \caption{LLM-based pair-wise evaluation results for recommender systems with similar AUC \label{fig:similar_models}}
\end{figure}

\begin{figure}[t]
    \centering
    \includegraphics[width=0.45\textwidth]{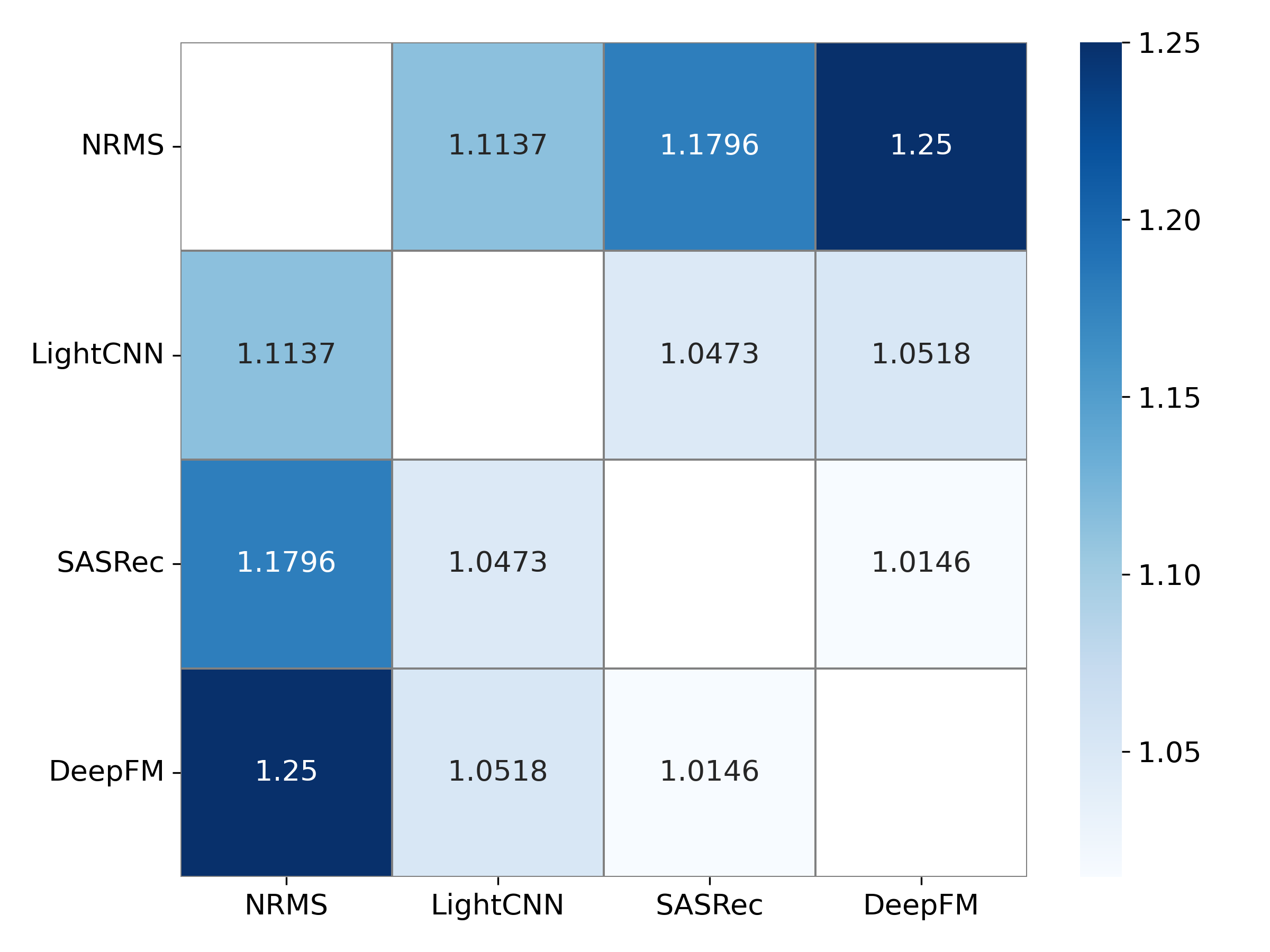}
    \caption{Values of $\mathcal{Q}$ in LLM-based pair-wise evaluation for recommender systems with similar AUC \label{fig:heatmap-q}}
\end{figure}

\textbf{Impact of different Models.}
In addition to recommendation models with significantly different recommendation accuracy (e.g., relative to FM), we also conducted pair-wise evaluations on recommendation models with similar AUC. 
The evaluation results (the percentage of wins) from GPT-4o are presented in Figure~\ref{fig:similar_models}.
We drew stacked bar-plots to show the results of the comparative evaluation of two recommendation models.
Additionally, we present the specific values of quantile $\mathcal{Q}$ for the pair-wise recommendation models in Figure~\ref{fig:heatmap-q}.
Figure~\ref{fig:similar_models} and Figure~\ref{fig:heatmap-q} show that for recommendation models with very similar AUC values, LLMs can effectively distinguish the quality of recommendations between them.
In other words, 
it implies that the LLM-based pair-wise evaluations can more effectively identify subtle differences in performance that offline metrics might not capture.

\finding{\textbf{Answer to RQ1}:The overall evaluation results obtained from the pair-wise evaluation method based on the LLM are consistent with the trends of offline metrics. Therefore, the pair-wise evaluation method based on the LLM can produce reliable evaluation results. Additionally, compared to offline metrics, LLM-based pairwise evaluations provide better discrimination.}

\subsection{RQ2: The Multiple-Aspect Performance of Evaluation}
The multiple-dimensions we designed in Section~\ref{sec:evaluation_dimensions} are intended to assist LLMs in deriving an overall evaluation result. Specifically, they guide the LLM to first consider the comparative evaluation results of these 6 sub-dimensions, and then synthesize them to arrive at an overall evaluation result. This step-by-step thought process is more conducive to obtaining reliable and accurate evaluation results.
However, we still need to verify the effectiveness of the sub-dimension evaluations.
We chose to validate the dimensions of inspiration and transparency.

\subsubsection{Inspiration}
\input{forms/rq2-inspiration-diversity}
Table~\ref{tab:rq2-inspiration} presents the pair-wise evaluation results provided by GPT-4o and DeepSeek-V2.5 in terms of inspiration aspect.
Table~\ref{tab:rq2-inspiration} shows that the evaluation trends provided by the LLM in terms of inspiration align well with the offline metric URD. Furthermore, in specific sub-dimensions (such as inspiration), the LLM's evaluation results also demonstrate better differentiation.

\subsubsection{Transparency}
The transparency metric is primarily used to measure whether the model has utilized sufficient and correct features to infer and predict the recommended item.
We assess whether the LLM can effectively evaluate the transparency metric by investigating whether it can perceive changes in the features used during training.
Specifically, we trained the DeepFM model using different features. One model was trained considering only the features $user\_id$ and $item\_id$. For the training of another model, we included more user features such as $age$, $gender$, and $occupation$.

\begin{figure}[t]
 \centering
 \subfigure[\textbf{Case 1}]{
    \begin{minipage}[b]{0.5\textwidth}
    \includegraphics[width=1\textwidth]{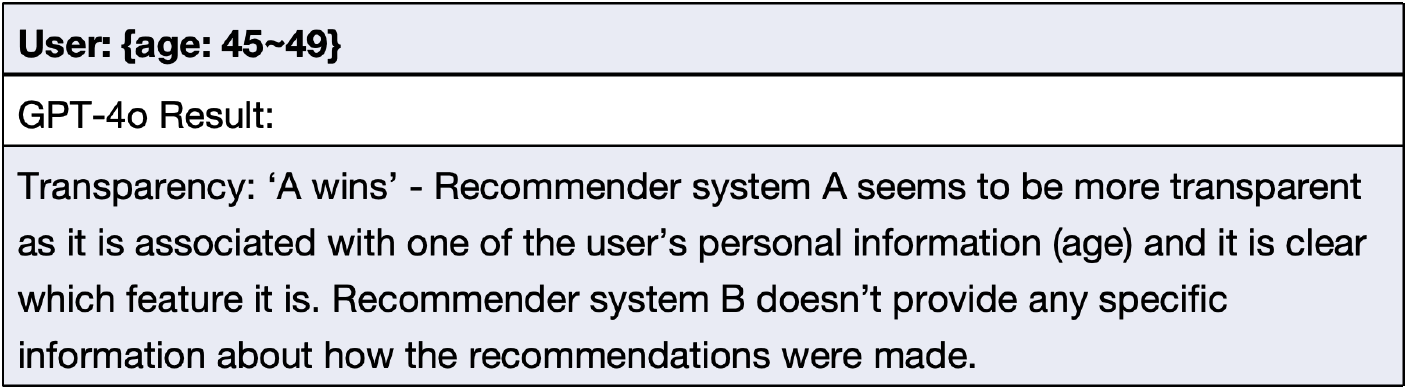}
    \end{minipage}
 }

 \subfigure[\textbf{Case 2}]{
    \begin{minipage}[b]{0.5\textwidth}
    \includegraphics[width=1\textwidth]{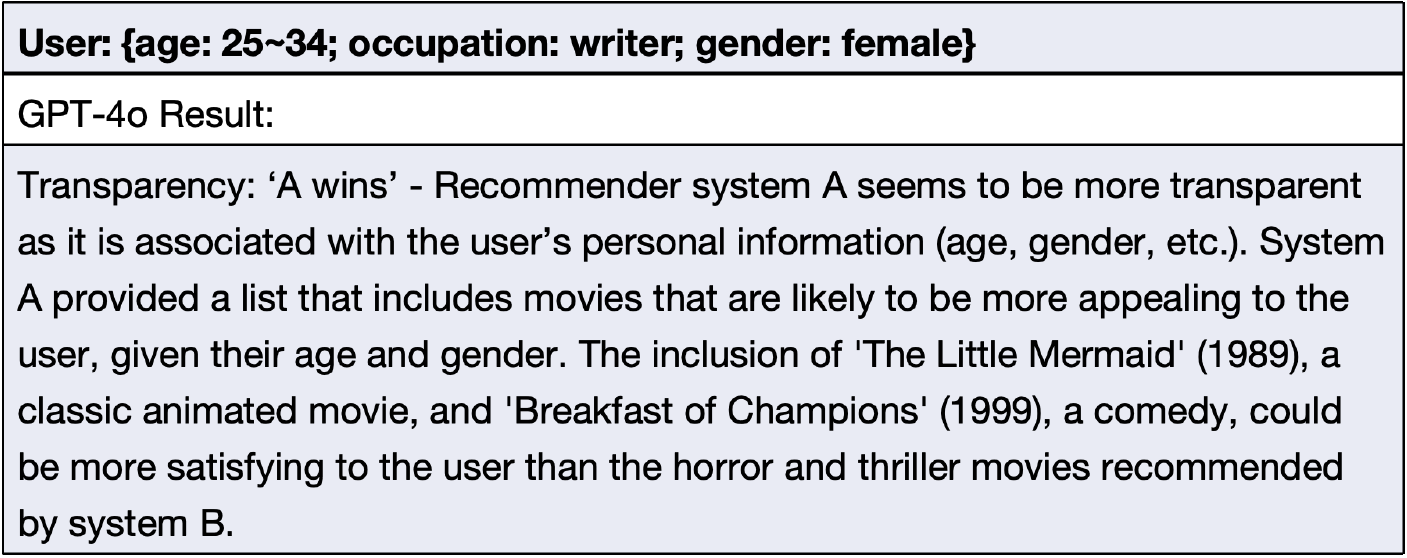}
    \end{minipage}
 }
 
 \caption{Two showcases of LLM-based pair-wise evaluation results for the transparency dimension}
 \label{fig:rq2-exp}
\end{figure}

In the Case 1, given a user aged 45-49, the recommendation results from recommender system B (a model trained only with $user\_id$ and $item\_id$) included movies in the category of "Children's". GPT-4o provided an evaluation on the transparency dimension, stating "A wins." The rationale for this assessment is that Recommender System A's recommendation results are more aligned with the user's age.
This example shows that LLMs are capable of distinguishing differences in recommendation results produced by recommendation models trained on different features. Furthermore, the LLM can pinpoint which specific feature is responsible for the observed differences.

In the Case 2, the user's specific characteristics are: age 25-34, occupation as a writer, and gender female.
However, Recommender System B recommended movies categorized as "Horror" to her, a genre that had not appeared in her historical viewing list.
From the results of GPT-4o, it is evident that LLMs can analyze and perceive that recommending movies categorized as "Horror" to this user is inappropriate based on her personal attributes.

\finding{\textbf{Answer to RQ2}: For sub-dimensions, the LLM-based pair-wise evaluation method can provide reasonable and discriminative assessment results.}







\subsection{Relative and Absolute Evaluation}

\begin{figure}[t]
    \centering
    \includegraphics[width=0.5\textwidth]{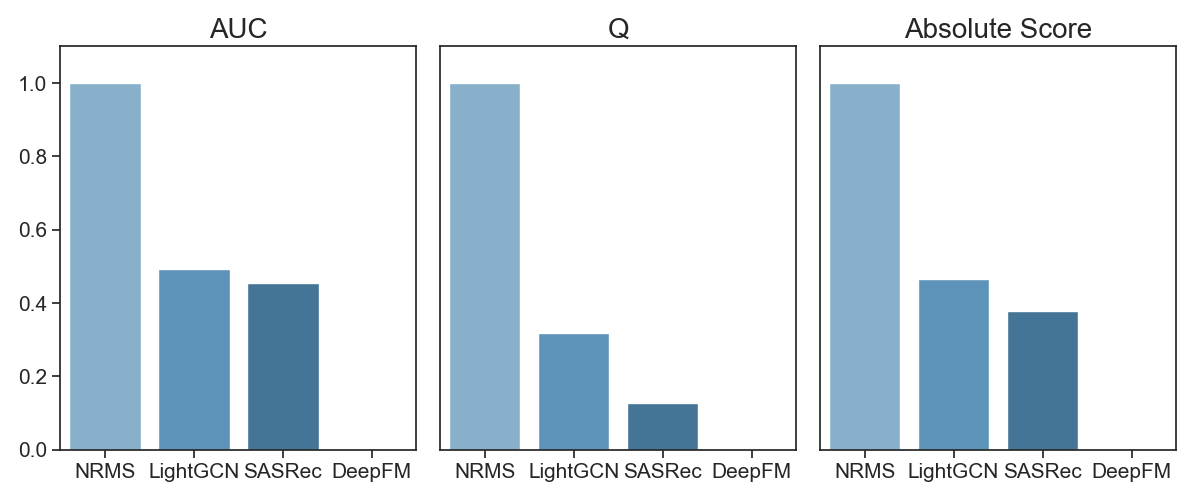}
    \caption{The differences among offline metric (AUC), LLM-based pair-wise evaluation reaults ($\mathcal{Q}$), and LLM-based absolute evaluation score \label{fig:absolut_score}}
\end{figure}

To investigate the differences in differentiation between relative and absolute evaluation using LLMs, we conducted a study in which we employed GPT-4o for the absolute evaluation of four recommender systems (i.e., NRMS, LightGCN, SASRec, and DeepFM) with similar AUC values trained on the MovieLens dataset. 
For each recommendation result list, we assigned a score between 0 and 1 using GPT-4o. 
In Figure~\ref{fig:absolut_score}, we present the normalized results for the offline metric (i.e., AUC), the pair-wise evaluation quantiles $\mathcal{Q}$, and the absolute evaluation scores for the four recommendation models: NRMS, LightGCN, SASRec, and DeepFM. 
Figure~\ref{fig:absolut_score} reveals that the LLM-based relative assessment provides better differentiation among these four models.
For example, for the two most closely matched models in terms of recommendation accuracy, LightGCN and SASRec, the quantiles $\mathcal{Q}$ derived from the LLM-based pair-wise evaluation still provide a better way to distinguish them, capturing the subtle differences between them.

%% file: forms/rq1-overview-movie.tex
\begin{table*}[t!]
\centering
\caption{The overall performance of LLM-based pair-wise evaluation \label{tab:rq1-movie}}
\begin{threeparttable}
\begin{tabular}{cc|ccccc|ccccc}
\toprule
\multirow{2}{*}{LLM}    &     \multirow{2}{*}{RS}        & \multicolumn{5}{c|}{\textbf{MovieLens}}                        & \multicolumn{5}{c}{\textbf{MIND}}                             \\
\cline{3-12}
                          &          & Win(\%)  & Tie(\%)  & Lose(\%) & Q      & Rank & Win(\%)  & Tie(\%)  & Lose(\%) & Q      & Rank \\ \hline
\multirow{4}{*}{GPT-4o}   & NRMS     & 61.1 & 14.7 & 24.2 & 1.9485 & 1  &71.9 & 11.0   & 17.1 & 2.9501 & 1  \\ 
                          & LightGCN & 62.8 & 8.4  & 28.8 & 1.9139 & 2 &60.5 & 20.9 & 18.6 & 2.0607 & 2   \\ 
                          & SASRec   & 60.3 & 10.7 & 29.0 & 1.7884 & 3 & 59.9 & 19.7 & 20.4 & 1.985  & 3   \\ 
                          & DeepFM   & 59.1 & 11.4 & 29.5 & 1.7237 & 4  & 55.0   & 15.2 & 29.8 & 1.56   & 4  \\ \hline
\multirow{4}{*}{DeepSeek-V2.5} & NRMS     & 64.7 & 12.6 & 22.7 & 2.1898 & 1  & 66.7 & 10.3 & 23.0   & 2.3123 & 1  \\ 
                          & LightGCN & 62.2 & 9.7  & 28.1 & 1.9021 & 2  & 62.6 & 15.3 & 22.1 & 2.0828 & 2  \\ 
                          & SASRec   & 61.3 & 9.2  & 29.5 & 1.8217 & 3  & 60.6 & 14.9 & 24.5 & 1.9162 & 3  \\ 
                          & DeepFM   & 64.7 & 12.6 & 22.7 & 1.7680 & 4  & 58.7 & 12.4 & 28.9 & 1.7215 & 4  \\ \hline
\multirow{4}{*}{Llama3.1-8B}    & NRMS     & 50.8 & 28.6 & 20.6 & 1.6138 & 3  & 54.8 & 27.1 & 18.1 & 1.8119 & 1  \\ 
                          & LightGCN &   50.5   &   37.9   &   11.6   &   1.7858     & 1  & 43.7 & 40.7 & 15.6 & 1.4991 & 3   \\ 
                          & SASRec   &  45.0    &   31.6   &   23.4   &   1.3927     & 4  & 50.3 & 35.6 & 14.1 & 1.7283 & 2   \\ 
                          & DeepFM   &  49.9    &   38.7   &   11.4   &   1.7684     &  2 & 45.1 & 34.9 & 20.0   & 1.4571 & 4   \\
\bottomrule
\end{tabular}
\begin{tablenotes}
 \footnotesize
\item[*]
  In the pair-wise evaluation, recommender system A $R_A$ is one of NRMS, LightGCN, SASRec, or DeepFM, while recommender system B $R_B$ is FM.
\end{tablenotes}

\end{threeparttable}
\end{table*}

%% file: forms/pearson.tex
\begin{table}[t!]
\centering
\caption{The Pearson Correlation between the LLM-based pair-wise evaluation results and the AUC \label{tab:pearson}}
\begin{adjustbox}{width=0.5\textwidth,center}
\begin{threeparttable}
\begin{tabular}{c|cc|cc}
\toprule
\multirow{2}{*}{LLM} & \multicolumn{2}{c|}{MovieLens} & \multicolumn{2}{c}{MIND}      \\ \cline{2-5}
                     & Correlation Coefficient & P-value & Correlation Coefficient & P-value \\ \hline
GPT-4o               & 0.8972              & 0.1027  & 0.9001              & 0.0998  \\
DeepSeek-V2.5        & 0.9436              & 0.0563  & 0.9530              & 0.0469  \\
Llama3.1-8B          & -0.2661             & 0.7338  & 0.7443              & 0.2556 \\
\bottomrule
\end{tabular}
\end{threeparttable}
\end{adjustbox}
\end{table}

%% file: forms/rq2-inspiration-diversity.tex
\begin{table}[t!]
\centering
\caption{The performance of LLM-based pair-wise evaluation in terms of inspiration \label{tab:rq2-inspiration}}
\begin{adjustbox}{width=0.5\textwidth,center}
\begin{threeparttable}
\begin{tabular}{c|c|cccccc}
\toprule
                          LLM &   RS  &   URD  & Win(\%)  & Tie(\%)  & Lose(\%) & Q      & Rank \\ \hline
\multirow{4}{*}{GPT-4o}   & SASRec & 0.1968   & 69.0 & 10.8 & 20.2 & 2.5741 &  1    \\ 
                          & LightGCN & 0.1962 &  60.1   &   4.9   &   35.0   &  1.6290      &  2    \\ 
                          & NRMS  & 0.1963 & 59.6 & 5.6 & 34.8 & 1.6138 &  3    \\ 
                          & DeepFM  & 0.1954 &   57.3   &   11.0   &   31.7   &  1.5995      &  4    \\ \hline
\multirow{4}{*}{DeepSeek-V2.5} & SASRec  & 0.1968   &  62.8    &    12.7  &   24.5   &    2.0295    &   1   \\ 
                          & LightGCN & 0.1962 &   58.9   &    20.5  &  20.6    &    1.9318    &  2    \\ 
                          & NRMS  & 0.1963 &   61.0   &   8.3   &   30.7   &     1.7769   &   3   \\ 
                          & DeepFM & 0.1954  &  56.9    &   11.1   &   32.0   &    1.5777    &  4    \\
\bottomrule
\end{tabular}
\begin{tablenotes}
 \footnotesize
\item[*]
  In the pair-wise evaluation, recommender system A $R_A$ is one of NRMS, LightGCN, SASRec, or DeepFM, while recommender system B $R_B$ is FM.
\end{tablenotes}
\end{threeparttable}
\end{adjustbox}
\end{table}

%% file: content/relatedWork.tex
\subsection{Recommender Systems Evaluation}

Evaluation is a well-established core part of the recommendation field.
Generally, it aims to measure how well a recommender system can produce a list of ranked items that match the user's preferences.
Currently, researchers typically use user clicks and ratings to quantify user preferences.
The traditional evaluation process is divided into three steps: sampling evaluation data, designing evaluation metrics, and implementing evaluation methods.

\textbf{Sampling.}
For evaluation data, Li et al.~\cite{li2020sampling} study the relationship between sampling and the global top-k Hit-Ratio (HR, or Recall).
Li et al. demonstrate theoretically and experimentally that sampling top-k Hit-Ratio provides an accurate approximation of its global exact counterpart and can consistently predict the correct winner.
Krichene et al. argue that the sampled metrics can be viewed as high-bias, low-variance estimators of the exact metrics~\cite{krichene2020sampled}.
Therefore, they correct the sampled metrics point-by-point by minimizing criteria that trade-off bias and variance.

\textbf{Metrics.}
Avazpour et al.~\cite{avazpour2014dimensions} made a comprehensive summary of the evaluation metrics of recommender systems from 16 dimensions, including correctness, coverage, diversity, trustworthiness, confindence, novelty, serendipity, utility, risk, robustness, learning rate, usability, scalability, stability, privacy and user preference. 
They also analyzed the positive and negative correlations between the various metrics.
To better study the concepts and metrics of recommendation evaluation, it is important to consider what constitutes user satisfaction~\cite{silveira2019good}. The ultimate goal of recommender system is to satisfy the interest of users.
Therefore, on this basis, Kim et al.~\cite{kim2021customer} designed a user satisfaction metric that comprehensively considered accuracy and diversity.
Zhao et al.~\cite{zhao2020revisiting} studied the recommendation for different domains, the difference and correlation between model effectiveness.
Maksai et al.~\cite{maksai2015predicting} propose a set of evaluation metric system for news recommender system.

\textbf{Evaluation Approaches.}
Evaluation methods can be divided into two categories, evaluation using collected data (i.e., offline test)~\cite{gilotte2018offline, mcinerney2020counterfactual, clarke2021evaluation} and evaluation using real user interaction~\cite{lu2021standing, vsafavrik2022repsys, zhang2020evaluating, bountouridis2019siren, luo2022mindsim}.
In addition to using data and calculating metrics in various dimensions, some researchers have introduced methods like preference graphs~\cite{clarke2021evaluation} and counterfactual evaluation~\cite{gilotte2018offline, mcinerney2020counterfactual} to better mine evaluation information.
A/B testing is a common test method to verify whether a change will have a significant impact on core metrics.
Usually, in industry, we split the traffic (i.e., experimental users) in equal proportions, then apply two sets of schemes, and finally compare the metric results.
However, this method of online a/b testing is difficult to implement in academic research. 
McInerney et al.~\cite{mcinerney2020counterfactual} propose a new counterfactual estimator using interaction causal relationships to reduce variance.
Interaction-based evaluation methods can be further subdivided into interaction evaluation methods with real users (i.e., online test)~\cite{lu2021standing, vsafavrik2022repsys} and user simulation~\cite{zhang2020evaluating, bountouridis2019siren, luo2022mindsim}.
Lu et al.~\cite{lu2021standing} propose a method to manually annotate user preferences for the evaluation of recommender systems.
Combining data visualization techniques, Šafařík et al.~\cite{vsafavrik2022repsys} proposed a set of highly interactive approaches for the evaluation of recommender systems.
However, real user interaction is usually difficult to obtain. The researchers propose a series of simulators of user behavior~\cite{zhang2020evaluating, bountouridis2019siren, luo2022mindsim}.
Luo et al.~\cite{luo2022mindsim} leverage reinforcement learning to simulate users' click decisions.

\subsection{LLM-based Evaluation}
Recent advances in LLMs have demonstrated impressive capabilities on a broad range of tasks~\cite{wang2024survey, xi2023rise, wu2024survey}.
Previous works have considered using LLMs for evaluation tasks~\cite{kocmi2023large, zhang2024large, wang2023rethinking, oosterhuis2024reliable}.
Kocmi et al.~\cite{kocmi2023large} propose using LLMs to evaluate the translation quality. They found that the effectiveness of the LLM-based evaluation is unexpectedly high, even surpassing all existing metric-based evaluation methods.
Zhang et al.~\cite{zhang2024large} proposed that certain zero-shot LLMs can achieve comparable or even better evaluation accuracy compared to traditional methods in the task of evaluating recommendation explanation quality. In addition, they claim that using the voting results of multiple LLMs can improve the accuracy of evaluations.
Wang et al.~\cite{wang2023rethinking} proposed utilizing the role-play capability of LLMs and using them as user simulators to evaluate conversational recommendation systems.
Recently, there have also been research works to evaluate the quality of other LLMs using LLMs~\cite{chang2024survey}.
Additionally, some studies have utilized LLMs for relevance annotation in an IR context~\cite{clarke20234, faggioli2023perspectives, macavaney2023one, thomas2024large}.

Different from them, 1) evaluating recommendation tasks involves a larger amount of information, requiring a comprehensive consideration of both user and item information. The complexity of user features and the dynamics of user preferences increase the difficulty of assessing the quality of recommendations. 2) In addition to accuracy, recommendation quality assessment involves more dimensions. Since recommender systems are user-facing software, the evaluation results tend to be more subjective, focusing on user experience. Such evaluation reports are more useful for the platform's development and more valuable for guiding improvements.